\documentclass[preprint2]{aastex}

\shorttitle{SFDM overview}
\shortauthors{T. Matos & V.H. Robles}

\begin{document}


\title{Scalar Field (Wave) Dark Matter}


\author{Tonatiuh Matos\altaffilmark{*} and Victor H. Robles\altaffilmark{**}}
\affil{Departamento de F\'isica,Centro de Investigaci\'on y de Estudios Avanzados del IPN, AP 14-740, 0700  D.F., M\'exico}


\altaffiltext{*}{Electronic address: \tt{tmatos@fis.cinvestav.mx}}
\altaffiltext{**}{Electronic address: \tt{vrobles@fis.cinvestav.mx}}

\begin{abstract}
Recent high-quality observations of dwarf and low surface brightness (LSB) galaxies have shown that their dark matter (DM) halos 
prefer flat central density profiles. On the other hand the standard cold dark matter model simulations predict a more cuspy behavior.
Feedback from star formation has been widely used to reconcile simulations with observations, this might be successful in field dwarf 
galaxies but its success in low mass galaxies remains uncertain. 
One model that have received much attention is the scalar field dark matter model. 
Here the dark matter is a self-interacting ultra light scalar field that forms a cosmological Bose-Einstein condensate,
a mass of $10^{-22}$eV/c$^2$ is consistent with flat density profiles in the centers of dwarf spheroidal galaxies, reduces the 
abundance of small halos, might account for the rotation curves even to large radii in spiral galaxies and has an early galaxy formation.
The next generation of telescopes will provide better constraints to the model that will help to distinguish 
this particular alternative to the standard model of cosmology shedding light into the nature of the mysterious dark matter. 
\end{abstract}

\keywords{dark matter: scalar field}

\section{INTRODUCTION}

The standard model of cosmology assumes the dark matter is cold and effectively collisionless, the  
galaxies are formed in a hierarchical way, and as they evolve they are subject to frequent collisions and interactions 
with nearby galaxies that determined the properties that we observed today. 

The standard model, also called cold dark matter (CDM) model, is remarkably successful to describe the large scale structure 
of the universe, as well as large scale observations. Nowadays, galactic observations are becoming more precise that 
it is possible to assess some of the predictions from the CDM model with more reliability. Moreover, the numerical simulations 
are rapidly reaching the required resolution to study the inner parts of dwarf galaxies, that is within $\sim$500pc. 
Increasing the resolution has revealed that some discrepancies between the observations and the theoretical expectations 
might require careful revision to our understanding. 

One of them is the longstanding core/cusp discussion, whether the central dark matter (DM) profiles in dwarfs and low surface brightness (LSB) 
galaxies are more core-like and rounder than the standard cold dark matter (CDM) model predicts (see \citet[]{eym09,blo10} for a review). 
The core profiles most frequently used in the literature and that best fit the observations are 
empirical \cite[]{bur95,kuz10}. Albeit useful to characterize properties of galaxies, it is desirable to 
find a theoretical framework capable to produce the cores, since CDM suggest central densities in small galaxies going as 
$\rho \sim r^{-1}$ at small $r$ \citep{nav10} whereas observations of LSB galaxies 
suggest a core-like behavior ($\rho \sim r^{-0.2}$) \cite[]{seheon11,rob12,blo01,kuz11,kuz11b,boy14}. 
The trend to reduce the inner logarithmic slopes invokes astrophyisical processes such as radiation wind, supernovae feedback,
etc. \citep{gov10,gov12,mer06,gra06}, although this seems possible in LSB galaxies, the question remains for the fainter 
galaxies where one supernovae could blow out most of the gas due to the shallower gravitational potential. 

Another discrepancy possibly related the overabundance of satellites \cite[]{kly99,moo99,mac12,gar14} 
is the Too-Big-to-Fail \cite[]{boy11,gar14b} issue. The latter results from the higher number of massive dark matter 
halos around Milky-Way like host with the most massive galaxies observed in our local neighborhood, assuming the most massive 
galaxies are in the most massive dark matter halos, there should be about $\sim 10$ more around systems with virial masses comparable 
to our Milky Way or M31 (Andromeda) \cite[]{gar14b}. There have been some possible solutions, most of them relying on tidal stripping 
in addition to supernovae feedback.

However, some of these discrepancies might also be solved assuming different properties for the 
dark matter, such as scalar field dark matter (SFDM) \cite[]{sin94,lee96,guz00,mat01}, 
strongly self-interacting DM \cite[]{spe00,vog12,roc13}, warm dark matter \citep{mac12}. 

It is of particular interest to us the SFDM alternative, here the mass of the field is assumed to be 
very small ($\sim 10^{-22}$eV/$c^2$) such that its de Broglie wavelength is of order $\sim$~kpc, relevant for galactic scales. 
The quantum behavior of the field has created much interest in the model due to its success to account for some discrepancies
mentioned above with dark matter properties only, for example, the small mass keeps the central density from 
increasing indefinitely due to the uncertainty principle in contrast to CDM simulations where supernova feedback is 
required\citep{gov10,gov12,pon12,sca13}.

\section{SFDM: Previous work}

The main idea in the scalar field dark matter model \citep{sin94,ji94,lee96,guz00,hu00,mat01} considers 
a self-interacting scalar field with a very small mass, typically of $\sim 10^{-22}$eV/$c^2$, such that the quantum 
mechanical uncertainty principle and the interactions prevent gravitational collapse in self-gravitating structures,
thus the halos are characterized with homogeneous densities (usually referred as a cores) in their centers, 
in general the core sizes depend on the values of the mass and the self-interacting parameters \citep{col86} (for a review 
see \cite{sua13,rin14}). From the particle physics point of view the most simple way to account
for a scalar field with this features is adding a Higgs-like term with a mass $\sim 10^{-22}$eV/$c^2$
to the standard model of particles \citep{mat14}.

Previous studies of the cosmological evolution of a scalar field with mass $m\sim$10$^{-22}$eV/$c^2$ 
have shown that the cosmological density evolution is reproduced and very similar to the one obtained from CDM \citep{mat01,cha11,sua11,mag12,sch14}, 
there is consistency with the acoustic peaks of the cosmic microwave background radiation \citep{mat01,rod10} and this small mass implies a 
sharp cut-off in the mass power spectrum for halo masses below $10^{8}$M$_{\odot}$ suppressing structure formation of low mass 
dark matter halos \citep{mat01,mar14,boz14,hu00}. Moreover, there is particular interest in finding 
equilibrium configurations of the system of equations that describe the field (Einstein-Klein-Gordon system) and of its 
weak field approximation (Schr\"{o}dinger-Poisson(SP) system), different authors have obtained solutions interpreted 
as boson stars or later as dark matter halos showing agreement with rotation curves in galaxies and 
velocity dispersion profiles in dwarf spheroidal 
galaxies \citep{lee96,boh07,rob12,rob13,lor12,lor14,med15,die14,guz14}.
So far the large and small scales observations are well described with the small mass and thus has been taken as a prefered value 
but the precise values of the mass and self-interaction parameters are still uncertain, tighter constraints can come  
from numerical simulations \citep{sch14} and modeling of large galaxy samples. 

Recently the idea of the scalar field has gained interest, given the uncertainty in the parameters 
the model has adopted different names in the literature depending on the regime that is under discussion, 
for instance, if the interactions are not present and the mass is $\sim 10^{-22}$eV/$c^2$ this limit 
was called fuzzy dark matter \citep{hu00} or more recently wave dark matter \citep{sch14}, 
another limit is when the SF self-interactions are described with a quartic term in the scalar field potential and 
dominate over the mass (quadratic) term, this was studied in \citep{goo00,sle12} and called repulsive dark matter or 
fluid dark matter by \citep{pee00}. 

Notice that for a scalar field mass of $\sim 10^{-22}$eV/$c^2$ the critical temperature of condensation for the 
field is T$_\mathrm{crit}\sim m^{-5/3}\sim$TeV, which is very high, if the temperature of the field is below its 
critical temperature it can form a cosmological Bose Einstein condensate, if it condenses it is called Bose-Einstein condensed(BEC) 
dark matter \citep{mat01,guz00,ber10,rob13,har11,cha11}. \citet{sik09} mentioned that axions could also form Bose-Einstein 
condensates even though their mass is larger than the previous preferred value, notice that the result was contested in \citet{dav13}, 
this suggest that the condensation process should be study in more detail to confirm it can remain as BEC dark matter. 
In \cite{ure09}, it was found that complex scalar field with $m<10^{-14}$eV/$c^2$ that decoupled being still 
relativistic will always form a cosmological Bose-Einstein condensate described by the ground state wave function, 
this does not preclude the existence of bosons with higher energy, particularly in dark matter halos. 

We see that the smallness of the boson mass is its characteristic property and cosmological condensation is a likely consequence.
The preferred mass of the scalar field dark matter points to be close to $\sim 10^{-22}$eV/$c^2$, consistent with the above constraint, 
although there are still uncertainties on the mass parameter, in order to avoid confusion with the known QCD axion, 
we find it useful and appropriate to name the scalar field dark matter candidate, from the above characteristics we can 
define it as a particle with mass $m<10^{-14}$eV/$c^2$, we name this DM candidate the \textit{psyon}.

It is worth emphasizing that despite the variety of names given to the model the main idea described above remains the same, 
it is the quantum properties that arise due to the small mass of the boson that characterize and distinguishes this paradigm,
analoguous to the standard cosmological model represented by the CDM paradigm whose preferred dark matter 
candidates are the WIMPs (weakly interacting massive particles), one being the neutralino, 
we see that for all the above regimes SFDM, Repulsive DM, Axion DM, or any other model assuming an ultra light bosonic particle
comprise a single class of paradigm, which we categorized as the \textit{Quantum Dark Matter} (QDM) paradigm. 
As pointed before, in the QDM paradigm the small mass of the dark matter boson leads to the possibility of forming cosmological 
condensates, even for axions which are non-thermally produced and have masses in $10^{-3}-10^{-6}$eV/$c^2$ \citep{sik09}, 
this is a characteristic property that distinguishes these dark matter candidates from WIMPs or neutrinos, 
namely, the existence of \textit{bosons in the condensed state}, or simply \textit{BICS}, thus the axion and psyon are 
BICS.

 \section{Scalar field dark matter halos}

There has been considerable work in finding numerical solutions to the non-interacting SFDM in the non-relativistic regime 
to model spherically symmetric haloes \citep{guz00,guz04,ure10,ber10,bray10,kau68}, 
and also for the self-interacting SFDM \citep{boh07,rob12,col86,rin12,bal98,goo00}, 
it is worth noting that as mentioned in \cite{guz04} for the weak field limit of the system that determines the evolution of a spherically symmetric 
scalar field, that is, the Einstein and Klein-Gordon equations, for a complex and a real scalar field the system reduces 
to the Schr\"{o}dinger-Poisson (SP) equations \citep{arb03}.
The contraints reported in \cite{li14}, obtained by imposing that the SF behaves cosmologically as pressureless matter (dust), 
imply that the interacting parameter would be extremely small for the typical mass of $\sim$10$^{-22}$eV/$c^2$, 
therefore we expect that solutions to the SP system with no interactions would behave qualitatively similar to those when 
self-interactions are included, as supported by the similarity in the solutions of the non-iteracting case and those with a 
small self-coupling found in other works \citep{bal98,col86,bri11}.

One characteristic feature of stationary solutions of the form $\psi(\mathbf{x},t)= e^{-iE_n t}\phi(r)$ 
for the SP system is the appearance of nodes in the spatial function $\phi(r)$, these nodes are associated to different energy 
states of the SF, the zero node solution corresponds to the ground state, one node to the first excited state, and so on. 
These excited states solutions fit rotation curves (RCs) of large galaxies up to the outermost measured data and can even reproduce 
the wiggles seen at large radii in high-resolution observations \citep{sin94,col86,rob13}. 
However, halos that are purely in a single excited state seem to be unstable when the number of particles is 
not conserved (finite perturbations) and decay to the ground state with different decay rates \citep{guz04,bal98},
though they are stable when the number of particles is conserved(infinitesimal perturbations).
The ground state solution is stable under finite perturbations and infinitesimal perturbations \citep{ber10,sei90},
but has difficulties to correctly fit the rotation curves in large galaxies because its associated RC has a fast keplerian behavior 
after reaching its maximum value, hence unable to remain flat enough at large radii.

One way to keep the flatness of the RC to large radii is to consider that bosons are not fully in one state, but instead coexist in 
different states within the dark halo, these multistate halos (MSHs) have been studied in some works \citep{ber10,ure10,mat07,rob13,rob13b,med15}.
The size of the MSH is determined by the most excited state that accurately fits the RC for large radii, 
excited states are distributed to larger radii than the ground state, and in contrast to the halo with single state 
there are MSHs that are stable under finite perturbations provided the ground state in the final halo configuration 
has enough mass to stabilize the coexisting state \citep{ure10,ber10}. 

Although there are still uncertainties in the stability of the MSHs, the appearance of bosons in excited states seems to 
be a straightforward consequences of quantum interference triggered by halo mergers as confirmed recently in \cite{sch14b}, 
and possibly the internal evolution of the halo. Moreover,
initial fluctuations that grow due to the cosmological expansion of the universe eventually separate from it and start 
collapsing due to its own gravity, at this time (known as turnaround) the halo can have a number of psyons that are in different states 
which depend on the the local environment. Depending on the number density of bosons populating the excited states we can have different 
fates for the halos \citep{rob15}. 

On the other hand, including rotation in the halo might be needed, in fact, \citep{guz14} have included rotation to axis-symmetric 
halos in the condensed state and show that it can lead to the flattening of the RCs, other works have also included rotation but in the
context of MSHs in asymmetric configurations \citep{bray12}, both studies suggest that rotation is a relevant ingredient in halo modeling, 
in fact it should be, in the end we observed rotation in galaxies embedded in dark halos.  
However, we require more detail studies to assess the goodness of the agreement with a large sample of galaxies, especially 
because there ara several surveys underway (e.g. GAIA, MANGA) that will provide precise data to test the viability of 
the standard and alternative dark matter models.

\section{Conclusions}

There are several DM models in the literature that are addressing some discrepancies found in the standard model of cosmology, 
one of them is the scalar field dark matter. The quantum properties of the field affect kpc scales due to the smalleness of the 
mass. The typical psyon of mass $m \sim 10^{-22}eV/c^2$ reproduces the cosmological evolution just like CDM, it reduces the 
halo abundance in the faint end of the halo mass function offering a possible solution to the unobserved excess of satellites, 
and the Heinsenberg uncertainty principle generates shallow central densities in dwarf halos contrary to the cuspy profiles found in 
CDM simulations. It is clear that the SFDM model worths further exploration, in particular, improving the contraints in the
mass and interaction parameters such that we can distinguish unambiguously between CDM and SFDM. 

\acknowledgments

This work was supported in part by DGAPA-UNAM grant IN115311 and by CONACyT Mexico grant 49865-E.

\end{document}